\documentclass[preprint,showpacs,preprintnumbers,amsmath,amssymb]{revtex4}

\usepackage{graphicx}
\usepackage{epstopdf}
\usepackage{xcolor}
\usepackage{amsmath}
\usepackage{amstext,bm}
\usepackage{dcolumn}
\usepackage{bm}

\newcommand\mse{National Laboratory of Solid State Microstructures and Department of Materials Science
                and Engineering, Nanjing University, Nanjing 210093, China }

\newcommand\phy{National Laboratory of Solid State Microstructures and Department of Physics,
                Nanjing University, Nanjing 210093, China}
\newcommand\cmq{Collaborative Innovation Center of Advanced Microstructures, Nanjing University, Nanjing , 210093, China.}

\begin{document}

\title{Significant low lattice thermal conductivity and potential high thermoelectric figure of merit in Na$_2$MgSn}
\author{Cong Wang$^{1}$}
\author{Y. B. Chen$^{2}$}
\author{Shu-Hua Yao$^{1}$}
\author{Jian Zhou$^{1,3}$}  \thanks{Corresponding author: zhoujian@nju.edu.cn}
\affiliation{
$^1$ \mse \\
$^2$ \phy \\
$^3$ \cmq
}

\date{\today}

\begin{abstract}
             Thermoelectric materials enables the harvest of waste heat and directly conversion into electricity.
  In search of high efficient thermoelectric materials,  low thermal conductivity of a material
  is essential and critical. Here, we have theoretically investigated the lattice thermal conductivity and
  thermoelectric properties of  layered   intermetallic  Na$_2$MgSn and Na$_2$MgPb based
  on the density functional theory and linearized  Boltzmann equation with the single-mode relaxation-time approximation.
  It is found that both materials exhibit very low and anisotropic intrinsic lattice thermal conductivity.
  Despite of the very low mass density and simple crystal structure of Na$_2$MgSn, its lattice thermal conductivities
 along $a$ and $c$ axes are  only 1.75 and 0.80 W/m$\cdot$K respectively at room temperatures.
 When Sn is replaced by the heavier element Pb, its lattice thermal conductivities decrease remarkably
 to 0.51 and 0.31 W/m$\cdot$K respectively  along $a$ and $c$ axes at room temperatures.
 We show that the low lattice thermal conductivities of  both materials are mainly due to their very
  short phonon lifetimes, which are roughly between 0.4 to 4.5 ps.
          Combined with previous experimental measurements,  the metallic Na$_2$MgPb can not be a good
 thermoelectric material. However, we predict  that the semiconducting
 Na$_2$MgSn is a potential room-temperature thermoelectric material with a considerable $ZT$ of 0.34 at 300 K.
  Our calculations not only imply that the intermetallic Na$_2$MgSn is a potential thermoelectric material,
  but also can motivate more  theoretical and experimental works on the thermoelectric
  researches in simple layered intermetallic compounds.

\end{abstract}

\maketitle

\section{INTRODUCTION }
Thermoelectrical (TE) materials, which can directly convert heat to electricity,  have been extensively studied
for last several decades since  they play an important role in the area of environmentally
friendly energy technology.~\cite{Snyder_NM, Zebarjadi_EES}
The conversion efficiency of TE materials is usually evaluated by a dimensionless figure of merit,
$ZT=S^2\sigma T/\kappa $, where $S$ , $\sigma$, $T$, and $\kappa$ are the Seebeck coefficient, electrical conductivity,
absolute temperature and thermal conductivity respectively. The thermal conductivity $\kappa$ of a material can be further
divided into two items: $\kappa = \kappa_e + \kappa_L$, where $\kappa_e$ and $\kappa_L$ are electronic and lattice
thermal conductivity respectively.
For practical applications, the $ZT$ of a TE material should be at least larger than 1.
To be competitive, a higher $ZT$ of 4 is needed.\cite{DiSalvo_sci}
However, these physical quantities ($S$, $\sigma$, and $\kappa$) in the same material can not be
tuned separately due to their internal relationship. In general, the electrical conductivity ($\sigma$) and
electronic thermal conductivity ($\kappa_e$) will increase with the increase of carrier concentration,
while the Seebeck coefficient ($S$) will decrease with it. \cite{Snyder_NM}
Therefore, most good TE materials are semiconductors,
such as IV-VI compounds PbTe ~\cite{Heremans_science} and SnSe.~\cite{SnSe_nature, SnSe_science}

On the other hand, the lattice thermal conductivity ($\kappa_L$) is not directly related to
the carrier concentration. It is an effective method  to tune the $ZT$ values
of materials by tuning their lattice thermal conductivities.
Typical materials with low lattice thermal conductivities are glasses. However, glasses are
bad TE materials because of their very low carrier concentration and mobility compared
with crystalline semiconductors. \cite{Snyder_NM}  Therefore, good TE materials require a
rather peculiar property in the same system: `phonon-glass electron-crystal'. ~\cite{slack} In other words,
low lattice thermal conductivity is a necessary condition for good TE materials, although not a sufficient one.

In the classic physics picture, the lattice thermal conductivity can be approximated by the
formula: $\kappa_L = \frac{1}{3} C_v v l = \frac{1}{3} C_v v^2  \tau  $ , where $C_v$,
$v$, $l$, and $\tau$ are the heat capacity, phonon velocity,  mean free path (MFP),
and  relaxation time. Furthermore, the phonon velocity is often simply replaced by
the sound velocity, which is proportional to $\sqrt{ B/\rho} $, where $B$ and $\rho$ are
the elastic modulus and mass density of a material. \cite{Toberer}
Accordingly, one method to design low lattice thermal conductivity materials is to search for
 high density materials due to their low sound velocities, such as Bi$_2$Te$_3$.
Besides, complex crystal structural materials also usually have low sound velocities,
such as Yb$_{14}$MnSb$_{11}$.~\cite{Yb14MnSb11}
The other way is to reduce the relaxation time by introduction of defects or
nano-structures to scatter phonons. \cite{Toberer} Of course, the intrinsic large anharmonic
effect (large gr\"uneisen parameters) in a material will also reduce the relaxation time
by phonon-phonon scattering. A distinct example is SnSe, in which the large anharmonicity
leads to its exceptional low lattice thermal conductivity.~\cite{SnSe_nature}
However, this is not intuitive without quantitative calculations.

In this work, we predict by first principles calculations that the layered intermetallic
Na$_2$MgSn and Na$_2$MgPb have very low and anisotropic intrinsic lattice thermal conductivities.
Both materials have a very simple layered structure (8 atoms in the unit cell) and a low mass density
(2.82 and 4.01 g/cm$^3$ for Na$_2$MgSn and Na$_2$MgPb,  respectively).
In particular, we propose that Na$_2$MgSn is a promising room-temperature TE material.
An intermetallic is a solid-state compound exhibiting metallic bonding, defined stoichiometry
and ordered crystal structure. It is a large material family, which have a wide various crystal
 structures, ranging from zero to three in dimensionality.
Despite of their metallic bonding, some intermetallics are semiconductors,
which is the precondition for their TE applications.
There are many works about the potential TE intermetallic materials such as
 Mg$_3$Sb$_2$ \cite{Mg3Sb2,Mg3Sb2_snyder,Mg3Sb2_Bhardwaj}, CaMgSi \cite{CaMgSi},
 MGa$_3$ (M=Fe, Ru, and Os) \cite{TMGa3}, YbAl$_3$ \cite{YbAl3},  Zn$_4$Sb$_3$ \cite{Zn4Sb3},
 Al$_2$Fe$_3$Si$_3$ \cite{AlFeSi},  MIn$_3$ (M=Ru and Ir)  \cite{MIn3}, M$_2$Ru (M=Al and Ga) \cite{MRu} and etc.
 In particular,  many half- and full-Heusler compounds, which are magnetic intermetallics, are found to be good
 TE materials.~\cite{heusler1,heusler2,heusler3,heusler4}

In 2012,  Yamada \textit{et. al.} reported the synthesis, crystal structure, and basic physical properties
of hexagonal intermetallic  Na$_2$MgSn.~\cite{NaMgSn} They found that polycrystalline Na$_2$MgSn is a small band gap
 semiconductor with a  large Seebeck coefficient  of +390 $\mu$V/K
 and an electrical resistivity of about 10 m$\Omega$\ cm at 300 K.
As a result, the power factor of Na$_2$MgSn is almost 40\% of that of Bi$_2$Te$_3$.~\cite{NaMgSn}
Two years later, the same group  synthesized the similar intermetallic Na$_2$MgPb, ~\cite{NaMgPb}
which is a metal with three different phases from 300 to 700 K.
In the experiment, the electrical resistivity of Na$_2$MgPb is much lower than that of
Na$_2$MgSn, which is only 0.4 m$\Omega$\ cm at 300 K.
From the preliminary experimental results, Na$_2$MgSn could be a potential TE material.
However, the thermal conductivity of Na$_2$MgSn is not studied yet in both experiments~\cite{NaMgSn, NaMgPb}
and the following theoretical work. ~\cite{wang_NaMgSn}

The rest of the paper is organized as follows. In section II,
we will give the computational details
about phonon and thermal conductivity calculations.
In section III, we will present our main results about phonon
dispersions, temperature dependent and accumulated lattice thermal conductivity,
group velocity, Gr\"uneisen parameter, phonon lifetime, and Seebeck coefficient of  
Na$_2$MgSn and Na$_2$MgPb. Some comparisons between different TE materials are also given.
Finally,  a short conclusion is present.

\section{ COMPUTATIONAL DETAILS }

 The structural properties of Na$_2$MgSn and Na$_2$MgPb are calculated by
 the Vienna ab-initio simulation package (VASP) ~\cite{vasp1,vasp2} based on the density functional theory.
 The projected augmented wave method ~\cite{paw1,paw2} and the
 generalized gradient approximation with the
 Perdew-Burke-Ernzerhof exchange-correlation functional ~\cite{pbe} are used.
 The plane-wave cutoff energy is set to be 350 eV. Both the internal atomic positions and the
 lattice constants are allowed to relax until the maximal residual Hellmann-Feynman forces
 on atoms are smaller than 0.001 eV/\AA.
 An $8 \times 8 \times 4$ k-mesh was used in the optimization.

 Both the second- and third-order interatomic force constants (IFCs)
 are calculated by the finite displacement method.
 The second-order IFCs in the harmonic approximation and the phonon dispersions
 of Na$_2$MgSn and Na$_2$MgPb are calculated by the Phonopy code. ~\cite{phonopy}
 The third-order (anharmonic) IFCs and  the lattice thermal conductivity are
 calculated by the Phono3py code.~\cite{phono3py}
 We use a $2\times 2 \times 2$ supercell (64 atoms) for the calculations of the
 second- and third-order IFCS in Na$_2$MgSn and Na$_2$MgPb. And a q-mesh of $20\times 20\times 10$
 is used for the calculation of lattice thermal conductivity by Phono3py code.

 We do not use the crude force constants approximation in the third-order IFCS, although
 we have checked that a cutoff distance of 4 \AA \, can already obtain a good thermal conductivity.
 We also checked a larger supercell of  $3 \times 3 \times 2$  in Na$_2$MgSn with a cutoff distance of 5 \AA,
 and we found that the thermal conductivity changed little.

 The Seebeck coefficients of Na$_2$MgSn and Na$_2$MgPb are calculated by
 the BoltzTraP2 program \cite{boltztrap2} with the Boltzmann transport theory.
 The electron eigenvalues in the whole Brillouin zone are calculated by the VASP code
 with the hybrid functional of Heyd-Scuseria-Ernzerhof (HSE06) \cite{hse06} and spin-orbit coupling.
A k-mesh of 20 $\times$ 20 $\times$ 10 is used in the calculations of Seebeck coefficient.

\section{ RESULTS AND DISCUSSIONS }

\subsection{Crystal Structure and Phonon Dispersions}

\begin{figure}
\includegraphics[width=0.8\textwidth]{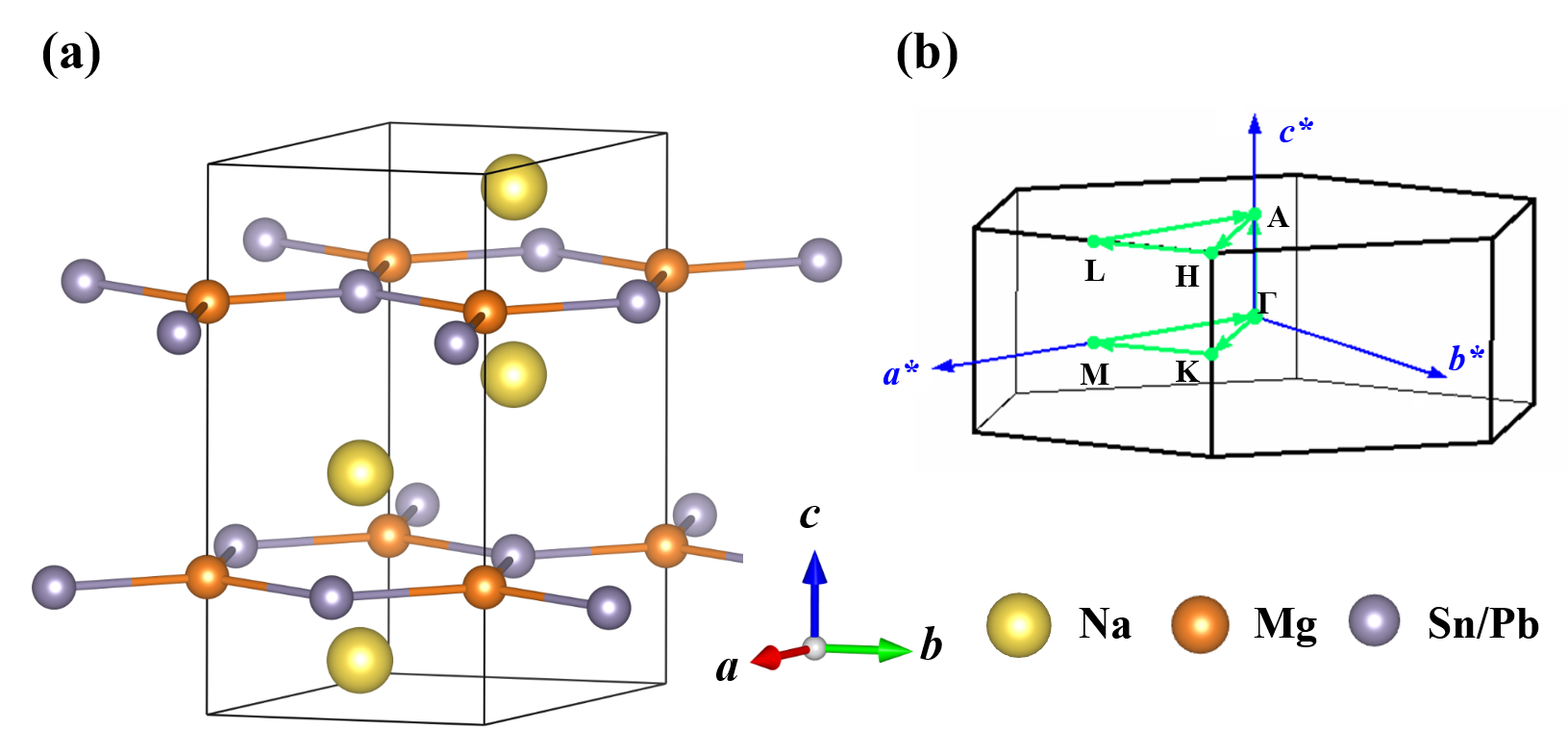}
\caption{\label{fig:structure} (Color online) ({\bf a}) Layered crystal structure and ({\bf b}) Brillouin zone and
high symmetry k-points of  hexagonal  Na$_2$MgSn and Na$_2$MgPb.  }
\end{figure}

\begin{table}
	\caption{ Calculated and experimental lattice constants of hexagonal Na$_2$MgSn and Na$_2$MgPb.
	}
	\begin{center}
		\begin{tabular}{c|c|c|c}
        \hline\hline
			Material    &  Method  &  $a$  (\AA)   & $c$  (\AA)     \\ \hline
						&  present work & 5.0825 & 10.1075   \\  \cline{2-4}
			Na$_2$MgSn  & exp.  (293 K) \footnote{ from reference~\cite{NaMgSn}  }  & 5.0486 & 10.0950        \\ \cline{2-4}
			            &  GGA-USP \footnote{from reference~\cite{wang_NaMgSn}  }   & 5.0085 &  10.1314    \\   \hline
			Na$_2$MgPb    &      present work   & 5.1415 & 10.1873  \\ \cline{2-4}
			            &   exp. (293 K) \footnote{from reference~\cite{NaMgPb} } & 5.1102 & 10.1714    \\  \hline \hline
		\end{tabular}
	\end{center}
\end{table}

As shown in Fig. 1(a), Na$_2$MgSn and Na$_2$MgPb
share the same hexagonal crystal structure with the space group of $P6_3/mmc$ (No. 194).
It is noted that  Na$_2$MgPb has three phases from 300 to 700 K.  \cite{NaMgPb}
However, from 300 to 500 K, Na$_2$MgPb and Na$_2$MgSn have the same hexagonal crystal structure. \cite{NaMgSn, NaMgPb}
Mg and Sn (or Pb) atoms lie in the same plane and form a two-dimensional (2D) honeycomb structure
stacking along the $c$ axis. Two layers of Na atoms are intercalated between the
adjacent Mg-Sn (or Mg-Pb) layers.
This is quite different from other alkali-intercalated layered materials, such as Na$_x$CoO$_2$  ~\cite{NaCoO}
 or Na$_x$RhO$_2$ ~\cite{zbb_NRO},  in which only one Na layer is intercalated between the adjacent CoO$_2$ or RhO$_2$ layers.
The Brillouin zone and high symmetry k-points of Na$_2$MgSn and Na$_2$MgPb are given in Fig. 1(b).

 The optimized lattice constants as well as the experimental ones of Na$_2$MgSn
 and Na$_2$MgPb are given in Table I.
 We  find that  the theoretical and experimental lattice constants are well consistent to each other,
 with the largest difference less than 1\%.  Our calculated lattice constant is also consistent with
 Wang's calculation by the generalized gradient approximation
 with the ultra-soft pseudopotentials (GGA-USP).  ~\cite{wang_NaMgSn}

\begin{figure}
	\includegraphics[width=0.9\textwidth]{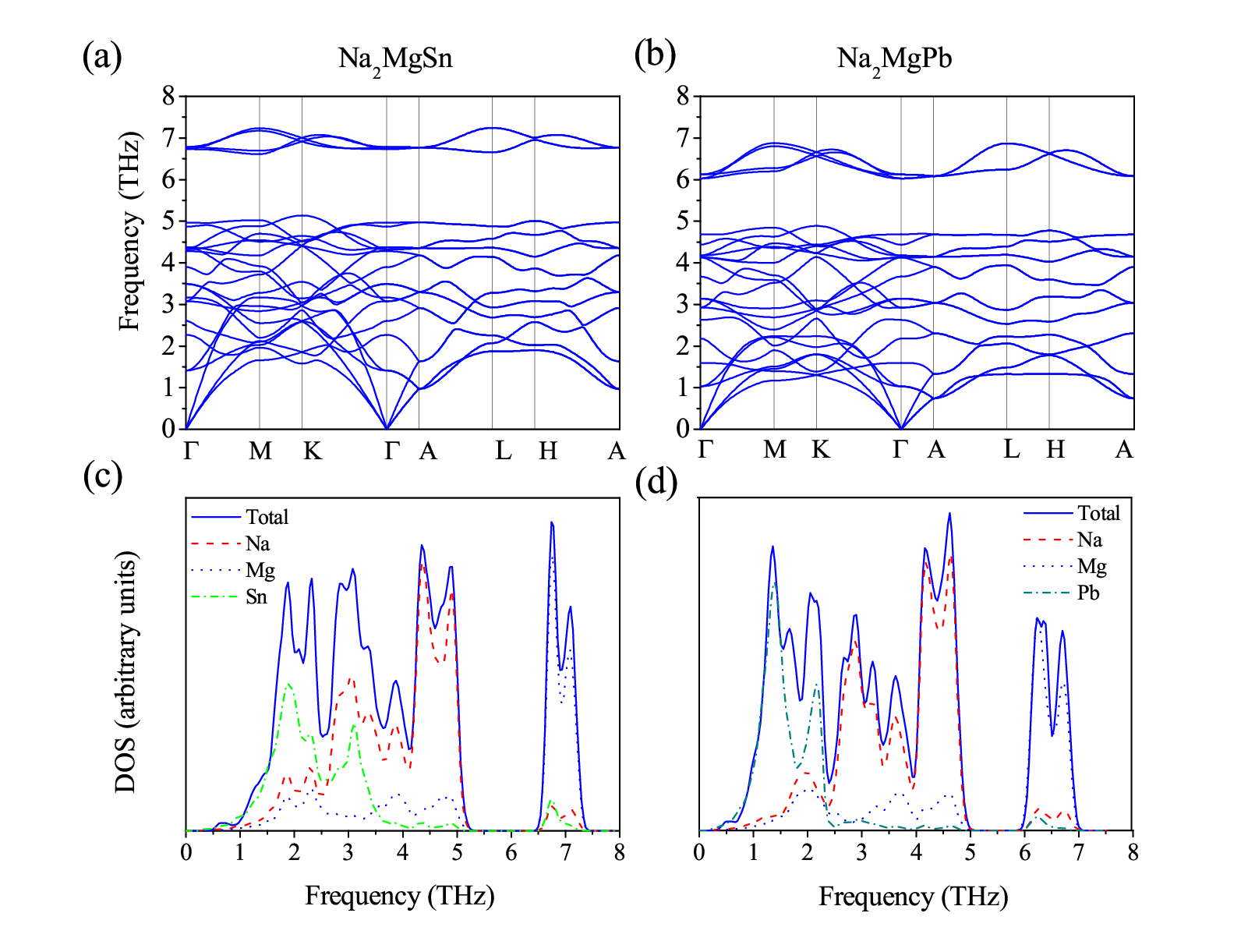}
	\caption{\label{fig:phonon} (Color online) Calculated  phonon dispersions and
                  density of states of  hexagonal Na$_2$MgSn (left column) and Na$_2$MgPb (right column).}
\end{figure}

Based on the optimized structures, the phonon dispersions and density of states (DOS)
of Na$_2$MgSn and Na$_2$MgPb are calculated by the Phonopy code,  given in Fig. 2.
It is obvious that  two materials show very similar phonon dispersions
 due to the same crystal structures. The highest frequency is about 7.5 and 7.0 THz
 for Na$_2$MgSn and Na$_2$MgPb respectively.  There is also a clear band gap
 from 5 to 6.5 THz for Na$_2$MgSn and from 5 to 6 THz for Na$_2$MgPb.

From the phonon DOS  in Fig. 2(c) and (d),
it is found that the high frequency  phonon modes above the band gap are mainly contributed
from the vibrations of Mg ions. This feature is same for Na$_2$MgSn and Na$_2$MgPb.
The main difference in phonon DOS between the two materials is the vibrations of Sn and Pb ions.
Due to the larger atomic mass of Pb ions, the vibrational frequencies of Pb ions are mainly below 2.5 THz in
Na$_2$MgPb, while the phonon modes of Sn ions extend from 0 to 4 THz in Na$_2$MgSn.
The vibrations of  Na ions are similar in both materials, which spread from 0 to 5 THz.
It is also noted that  the mid-frequency phonon modes from 2.5 to 5 THz in Na$_2$MgPb
are mainly contributed from the vibrations of Na ions. However, for the phonon modes in the same frequency range
in Na$_2$MgSn, there are also significant contribution from the vibrations of Sn ions.
In other words,  the vibrations of Na, Mg, and Pb ions in Na$_2$MgPb are well separated in different frequency regions,
while there is a  relatively large overlap  in Na$_2$MgSn.

\subsection{Lattice Thermal Conductivities}

\begin{figure}
	\includegraphics[width=0.7\textwidth]{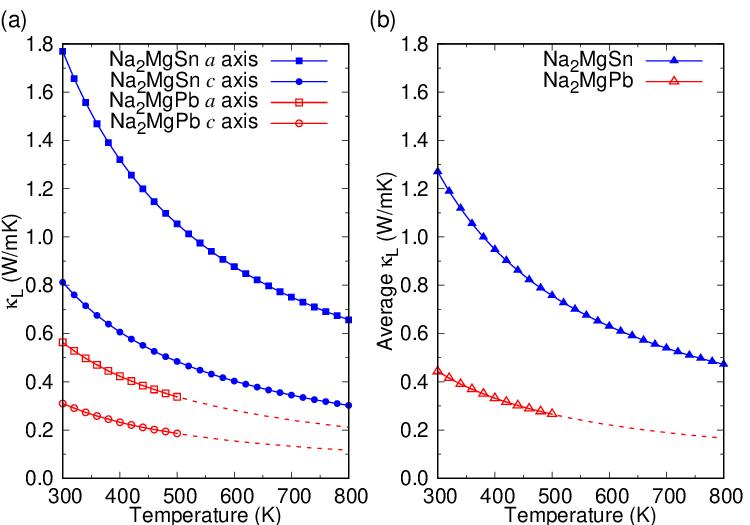}
	\caption{\label{fig:kappa} (Color online)   ({\bf a})   Calculated lattice thermal conductivities and  ({\bf b})  their average values
                               of hexagonal Na$_2$MgSn and Na$_2$MgPb from 300 to 800 K. The red dashed lines  mean
                               that the high temperature lattice thermal conductivities are calculated based on a low temperature crystal structure
                               of Na$_2$MgPb.  }
\end{figure}

\begin{table}
	\caption{ Calculated lattice thermal conductivities $\kappa_a$ and $\kappa_c$ along $a$ and $c$ axes
         and their average value $\bar{\kappa}$  of hexagonal Na$_2$MgSn and Na$_2$MgPb  at 300 K. The unit is W/m$\cdot$K.}
\begin{center}
\begin{tabular}{c|c|c|c}
  \hline\hline
  Material & $\kappa_a$ & $\kappa_c$ & $\bar{\kappa}$  \\ \hline
 Na$_2$MgSn  & 1.75  & 0.80  & 1.27   \\ \hline
 Na$_2$MgPb & 0.51 & 0.31  & 0.44 \\
  \hline\hline
\end{tabular}
\end{center}
\end{table}

Based on the harmonic and anharmonic IFCs, we have calculated the lattice thermal conductivity ($\kappa_L$)
by using the Phono3py code.
Fig. 3(a) shows the temperature-dependent thermal lattice conductivities along $a$ and $c$ axes
of Na$_2$MgSn and Na$_2$MgPb, while the average ones are also given in Fig. 3(b).
It is quite surprising to found that the $\kappa_L$ of the two intermetallics are
very low.  As shown in Fig. 3 and Table II, the $\kappa_L$ of Na$_2$MgSn is only 1.75 and 0.80 W/m$\cdot$K
along $a$ and $c$ axes at 300 K, while it is even much lower for  Na$_2$MgPb,
which is 0.51 and 0.31 W/m$\cdot$K  along  $a$ and $c$ axes at the same temperature.
Specifically, the  $\kappa_L$   along the $c$ axis in Na$_2$MgPb even approaches the predicted
amorphous limit (0.25 W/m$\cdot$K), \cite{Cahill_PRB} which is extremely low for crystalline solids.
The lattice thermal conductivity of  Na$_2$MgSn  is comparable with
the typical good TE materials, which will be  discussed  later.
While the lattice thermal conductivity of Na$_2$MgPb is even smaller than that of the
recently found best TE material SnSe,  which is 0.8, 2.0 and 1.7 W/m$\cdot$K
along $a$, $b$, and $c$ axes at 300 K from the first-principles calculations. \cite{Guo_PRB}
It is noted that the lattice thermal conductivity of Na$_2$MgPb above 500 K is meaningless
 since Na$_2$MgPb has different crystal structures above that temperature.
However, since we mainly focus on the room temperature behavior of Na$_2$MgPb,
it is not necessary to study its high temperature thermal conductivity.

Furthermore, it is obvious that Na$_2$MgSn and  Na$_2$MgPb both show an anisotropic lattice thermal conductivities
due to their layered crystal structures.  However, the ratio of thermal conductivities  between the $a$ and $c$
directions in  both materials is smaller than 2. The small anisotropy suggests that easily-formed texture
structures in layered compounds has not much effect on thermoelectric performance of these compounds.

We also calculated the average lattice thermal conductivity
$\bar{\kappa} $, defined by the formula $ 3/\bar{\kappa} = 2/\kappa_a  + 1/\kappa_c $,
shown in Fig. 3(b) and Table II.
The average $\bar{\kappa} $  for Na$_2$MgSn and Na$_2$MgPb are 1.27  and 0.44 W/m$\cdot$K at 300 K respectively.

We then  estimate the electronic thermal conductivity ($\kappa_e$) by the Wiedemann-Franz law:
$\kappa_e = LT\sigma$,
where $L$ is the Lorenz number (2.44 $\times$ 10$^{-8}$  W$\mathrm{\Omega}$K$^{-2}$),
$\sigma$ is the electrical conductivity, and $T$ is the absolute temperature.
According to previous experiments~\cite{NaMgSn, NaMgPb}, we can obtain the electrical
resistivities ($\rho$) of Na$_2$MgSn and Na$_2$MgPb are about 10 and 0.4 m$\Omega$\ cm at 300 K, respectively.
Therefore,  the  electronic thermal conductivity
of Na$_2$MgSn and Na$_2$MgPb are estimated to be 0.073 and 1.83 W/m$\cdot$K, respectively.
For the semiconducting Na$_2$MgSn, its electronic thermal conductivity is much smaller than the lattice one.
 While for the metallic Na$_2$MgPb,  the electron contributes more thermal conductivity than the lattice one.
 Of course, it has to be noted that the experimental samples are both polycrystals.
 No single crystal sample has been reported yet.

The sound velocity is also calculated by the slopes of  three acoustic phonon branches near the $\Gamma$ point.
For each directions, the sound velocity is averaged on the two transversal acoustic modes, (TA1 and TA2),
and one longitudinal acoustic mode (LA) by the formula:
$ {3}/{{v}_x^3} = 1/v_{x,\mathrm{TA1}}^3 + 1/v_{x,\mathrm{TA2}}^3  + 1/v_{x,\mathrm{LA}}^3 $,
where $x$ means the $a$ and $c$ axes.
In Table II, we can see that in $a$ and $c$ axes, the sound velocities of Na$_2$MgSn are both higher than those of Na$_2$MgPb,
which can  explain why Na$_2$MgSn has higher lattice thermal conductivities.
For both materials, the sound velocity along $a$ axis is higher than that along $c$ axis,
which can explain the anisotropic  lattice thermal conductivities in  two materials.

\begin{table}
	\caption{ Calculated sound velocities along $a$ and $c$ axes of Na$_2$MgSn and Na$_2$MgPb. The unit is km/s.}
	\begin{center}
		\begin{tabular}{c|c|c}
			\hline\hline
			Material & $ v_a$ &  $v_c$  \\ \hline
			Na$_2$MgSn  & 2.75  & 2.41    \\ \hline
			Na$_2$MgPb & 2.16 & 1.90 \\
			\hline\hline
		\end{tabular}
	\end{center}
\end{table}

\begin{figure}
	\includegraphics[width=0.8\textwidth]{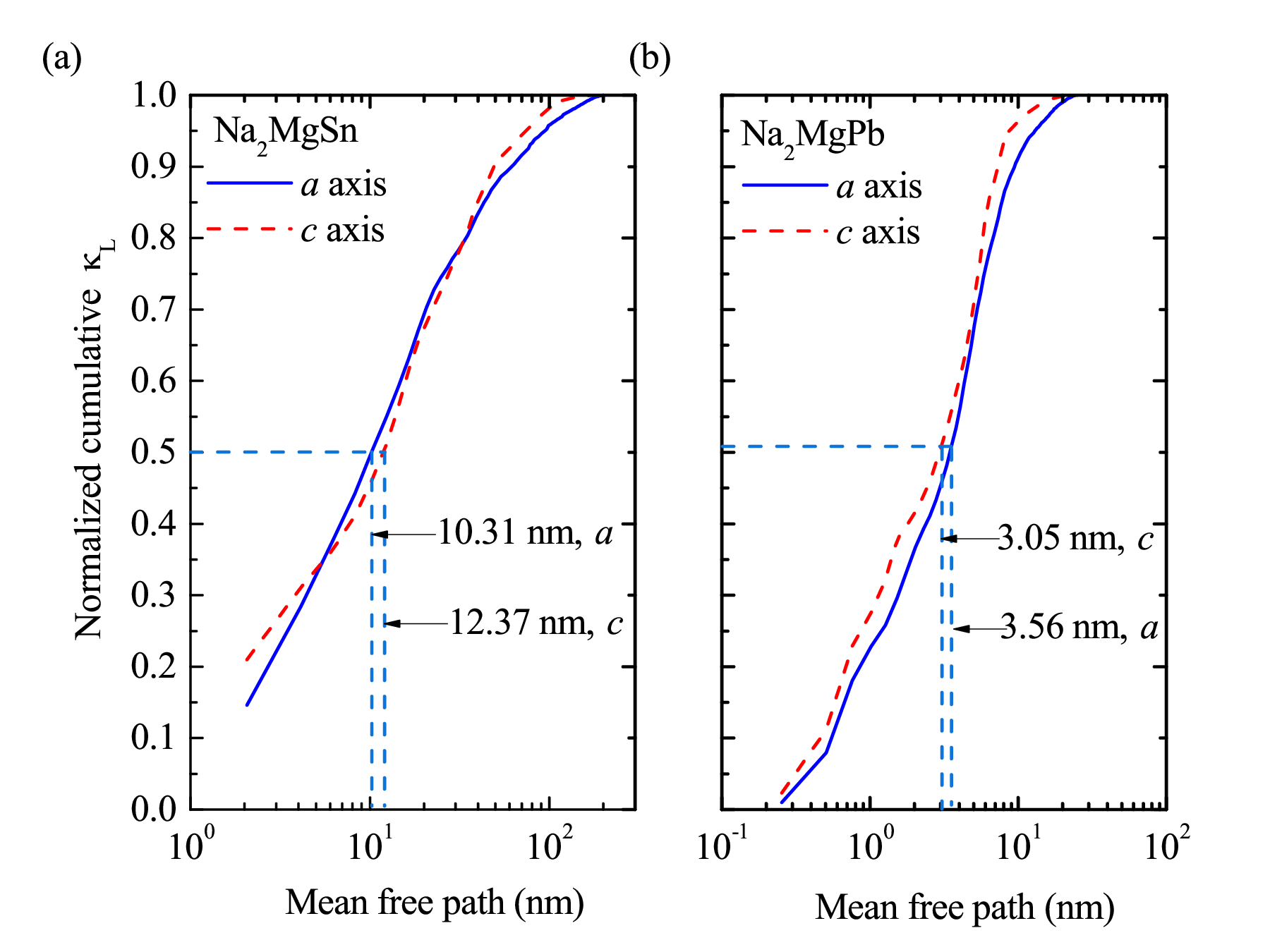}
	\caption{\label{fig:kappa} (Color online)  Normalized directional accumulated lattice thermal conductivities
                           of hexagonal ({\bf a})  Na$_2$MgSn and ({\bf b})  Na$_2$MgPb at 300 K. }
\end{figure}

We further plot the directional cumulative lattice thermal conductivity with respect to the
phonon MFP in Na$_2$MgSn and Na$_2$MgPb at 300 K in Fig. 4.
It is found that the $\kappa_L$ of Na$_2$MgSn is mainly dominated by the phonons whose MFPs are
less than 100 nm. However, for Na$_2$MgPb, the $\kappa_L$
is mostly contributed by the phonons whose MFPs are less than 20 nm.
This indicates that the phonon MFP in Na$_2$MgPb is much shorter than that in Na$_2$MgSn,
leading to a much lower $\kappa_L$ in Na$_2$MgPb than that in Na$_2$MgSn.

We  also give the representative MFP (rMFP) for the two materials in Fig. 4.
This parameter (rMFP) means that all the phonons whose MFP is shorter than the rMFP will contribute half to the total thermal conductivity.
It is clear to see that the  rMFP of Na$_2$MgPb is much shorter than that of Na$_2$MgSn.
 For Na$_2$MgSn, the rMFP along the $a$ and $c$ axes
are 10.31 and 12.37 nm respectively, while they are only 3.56 and 3.05 nm along the same axes for Na$_2$MgPb.
The rMFP of Na$_2$MgPb is  even  a little shorter than the ones of SnSe, which are  4.1, 4.9, and 5.6 nm
along $a$, $b$, and $c$ axes at 300 K from the first-principles calculations.  \cite{Guo_PRB}

 Due to their different MFP, we suggest that for Na$_2$MgSn,
 structural engineerings, such as nano-structuring or polycrystalline structures,
 can be taken to reduce the thermal conductivity effectively in experiments,
 while it would be much ineffective  for Na$_2$MgPb.

\subsection{Group velocity and phonon anharmonicity}

\begin{figure}
	\includegraphics[width=0.8\textwidth]{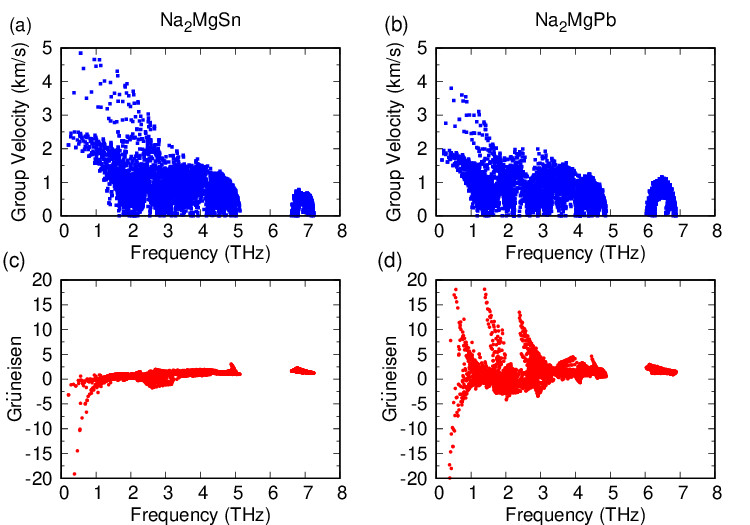}
	\caption{\label{fig:kappa} (Color online)   Calculated group velocities and mode-dependent
             Gr\"uneisen parameters of Na$_2$MgSn (left column)  and Na$_2$MgPb (right column) .}
\end{figure}

In order to further understand the intrinsic  thermal lattice conductivity
and their difference in Na$_2$MgSn and Na$_2$MgPb, we also calculated their frequency
dependent phonon group velocities,  mode-dependent  Gr\"uneisen parameter, and  phonon lifetimes.

In Fig. 5 (a) and (b),  we have given the magnitude of  frequency dependent phonon group velocities
 of Na$_2$MgSn and Na$_2$MgPb.
We can see that the group velocities of  Na$_2$MgPb are only slightly smaller than those of  Na$_2$MgSn,
which is difficult to explain the significant difference of $\kappa_L$  of the two materials.
Therefore,  we need to further study their anharmonic effect:  the Gr\"uneisen parameters.
In general, larger Gr\"uneisen parameter means larger anharmonicity of the material
and lower lattice thermal conductivity.  In Fig. 5 (c) and (d), the magnitude of Gr\"uneisen parameters for Na$_2$MgPb is
obviously much larger than that of Na$_2$MgSn, which means  that Na$_2$MgPb has a much stronger phonon anharmonicity.
In other words, stronger phonon scattering leads to lower lattice thermal conductivity in  Na$_2$MgPb
than that in Na$_2$MgSn.

The frequency-dependent phonon lifetimes of Na$_2$MgSn and Na$_2$MgPb are also
calculated by the Phono3py code from third-order anharmonic IFCs, plotted in Fig. 6.
The color bar in Fig. 6 represent the density of phonon modes.
In general, we found that the phonon lifetimes of both materials are very short, roughly ranging
from 0.4 to 4.5 ps , which are much smaller than those of SnSe (from 0 to 30 ps). \cite{Guo_PRB}
In detail, we can see that in the low frequency region below 2.5 THz, the phonon lifetimes of both materials
show a large broad distribution with a maximal value about 4.5 ps.
The lifetimes of the high frequency phonons (above the band gap) of the two materials are also similar with a
narrow distribution from 0.5 to 1.0 ps.
The major differences between the two materials are from the mid-frequency phonons, i.e.
from 2.5 to 5 THz. In this region, the phonons of Na$_2$MgSn have a relatively broader distribution from 0 to 1.5 ps,
while in Na$_2$MgPb, the phonons have much smaller lifetimes, located from 0.4 to 0.8 ps.
Therefore, we  conclude that the phonon modes  between 2.5 to 5 THz scatter
much stronger in Na$_2$MgPb than those in Na$_2$MgSn, which is the main reason
for the lower lattice thermal conductivity in Na$_2$MgPb than that of Na$_2$MgSn.

\begin{figure}
	\includegraphics[width=0.8\textwidth]{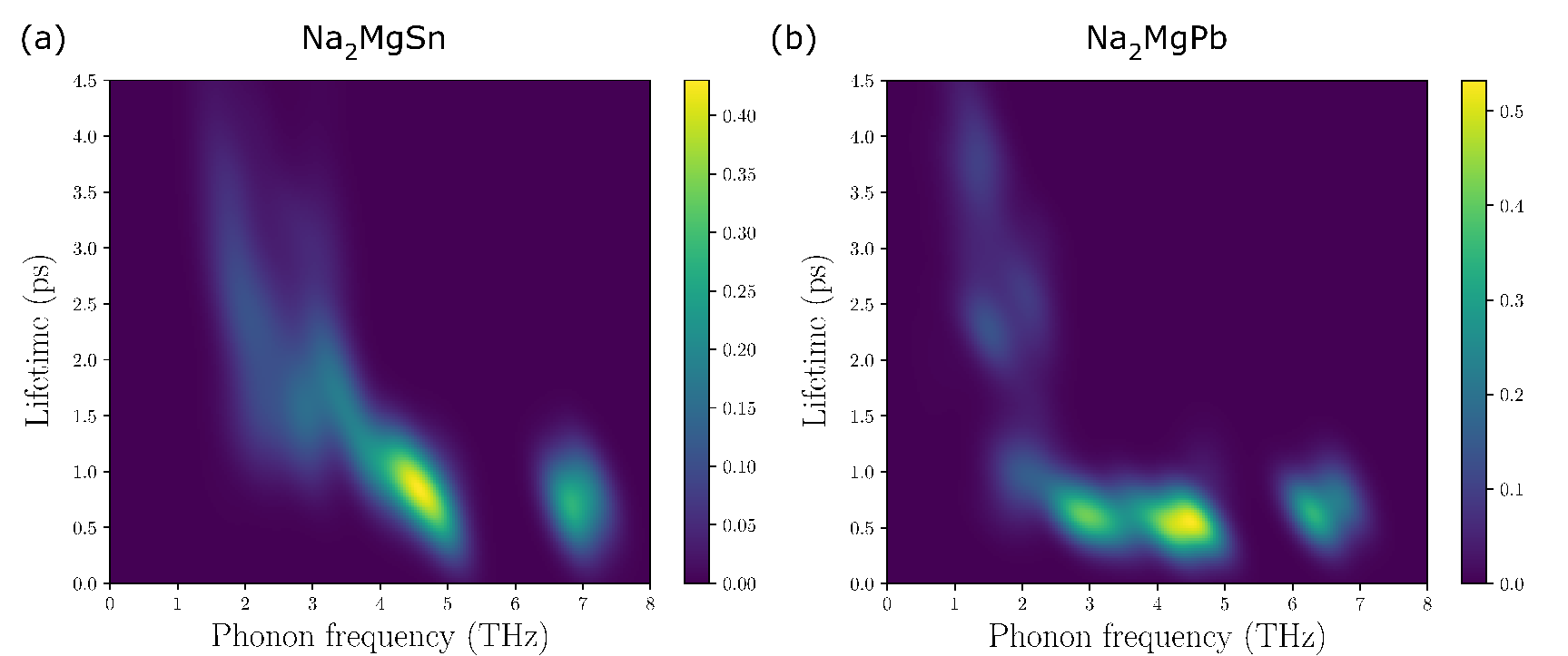}
	\caption{\label{fig:kappa} (Color online)  Calculated phonon lifetimes of   ({\bf a})  Na$_2$MgSn and  ({\bf b})   Na$_2$MgPb
         at 300 K.  The color in the firgure represents the phonon density. Brighter color means a higher phonon density.}
\end{figure}

\subsection{Seebeck coefficient}

 In order to estimate the $ZT$ of  Na$_2$MgPb, we have also calculated the Seebeck coefficient
 of two materials, shown in Fig. 7.  The electron band structure calculations indicate that
 Na$_2$MgSn is a small gap (about 0.18 eV)  semiconductor and Na$_2$MgPb is a semi-metal.
 In Fig. 7(a), the maximal absolute Seebeck coefficient of Na$_2$MgSn at 300 K is about 250
 $\mu$V/K, which is lower than the experimental value (390 $\mu$V/K). This is possibly because that
 the experimental Na$_2$MgSn is a polycrystalline sample.
 In Fig. 7(b),  it is natural to find  that  the Seebeck coefficient of  metallic Na$_2$MgPb is quite low.
 At the Fermi energy, its Seebeck coefficient  is about -22 $\mu$V/K.

\begin{figure}
	\includegraphics[width=0.8\textwidth]{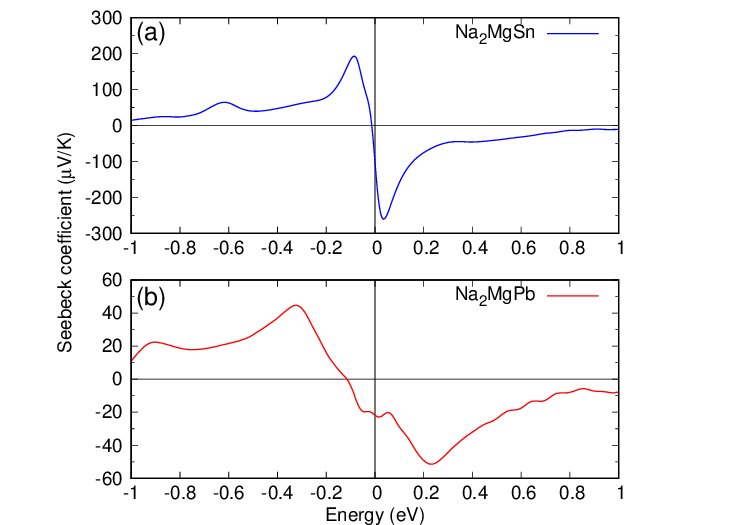}
	\caption{\label{fig:seebeck} (Color online)  Calculated Seebeck coefficient  of  ({\bf a})  Na$_2$MgSn and  ({\bf b})  Na$_2$MgPb
         at 300 K. The Fermi energy is set to 0. }
\end{figure}

\subsection{Discussion}

\begin{table}
	\caption{  Comparison of theoretical lattice thermal conductivity ($\kappa_L$), mass density ($\rho$),
         bulk modulus ($B$), sound velocity ($v$), and rMFP, and maximal phonon lifetime ($\tau$)
           of  Na$_2$MgSn and Na$_2$MgPb at 300 K with other TE materials.	 All physical quantities
           except the mass density are theoretical ones.  The data of Na$_2$MgSn and Na$_2$MgPb is from the present work.
          The (a), (b), and (c) in the table indicates the $a$, $b$, and $c$ axes respectively.}
	\begin{center}
		\begin{tabular}{c|c|c|c|c|c|c}
        \hline\hline
			Material& $\kappa_L$ (W/m$\cdot$K)  &   $\rho$  (g/cm$^3$) & $B$ (GPa)    & $v$ (km/s) & rMFP (nm) &$\tau$ (ps) \\ \hline
			Na$_2$MgSn 	&  1.75(a)/0.80(c)  & 2.82 \cite{NaMgSn}  & 30.6 & 2.75(a)/2.41(c)  & 10.31(a)/12.37(c) &$\sim$ 4.5  \\  \cline{2-7} \hline
			Na$_2$MgPb  &  0.51(a)/0.31(c)  & 4.01 \cite{NaMgPb} &  27.0  & 2.16(a)/1.90(c)  & 3.56(a)/3.05(c) &$\sim$ 4.5   \\  \cline{2-7} \hline
             SnSe  \cite{Guo_PRB}        & 0.8(a)/2.0(b)/1.7(c)   &  $\sim$6.1 \cite{SnSedensity}  & 39.4 &   0.40(a)/0.63(b)/0.58(c)
                                                    & 	5.6(a)/4.9(b)/4.1(c)    &  $\sim$ 30 \\  \cline{2-6}  \hline
            SnS \cite{Guo_PRB}          &  0.9(a)/2.3(b)/1.6(c)  &  5.1 \cite{SnSdensity}  & 41.6  & 0.44(a)/0.71(b)/0.60(c)
                                                   & 	4.3(a)/5.2(b)/4.0(c) &$\sim$ 30  \\  \cline{2-6} \hline
            Bi$_2$Te$_3$   & 1.2(a)/0.4(c) \cite{BiTe3}  &  7.88 \cite{Bi2Te3pccp}  & 36.4 \cite{Bi2Te3pccp}   & 1.68(a)/1.79(c) \cite{BiTe4}& 1.5(a) \cite{BiTe3}   \\  \cline{2-6}  \hline
            PbTe  &  2.1 \cite{PbTe1}     &  8.24 \cite{PbTe3}   & 45.51 \cite{PbTe4} & 1.98 \cite{PbTe4}  & 	6 \cite{PbTe1}  & $\sim$ 100  \cite{PbTe1} \\  \cline{2-6}
                          \hline\hline
		\end{tabular}
	\end{center}
\end{table}

Now, we compare some theoretical physical properties of Na$_2$MgSn and Na$_2$MgPb with other
well-known TE materials, such as SnSe, SnS, Bi$_2$Te$_3$, and PbTe, shown in Table IV.
It is found that  Na$_2$MgSn and Na$_2$MgPb have comparable lattice thermal
conductivities as those TE materials. The main difference
is that  Na$_2$MgSn and Na$_2$MgPb have a much smaller mass density, a smaller bulk modulus,
and a relatively  higher sound velocity. In particular, the mass density of Na$_2$MgSn is  less than
half of the SnSe's, while they almost have  the same lattice thermal conductivities.
This is a quite unique behavior in low $\kappa$ materials.
In spite of the very low mass density and high sound velocity, Na$_2$MgSn and Na$_2$MgPb show
 a very low lattice thermal conductivity due to their large anharmonicity.
 From Table IV, we can see that the maximal lifetime of Na$_2$MgSn and Na$_2$MgPb
 is much smaller than those of  SnSe, SnS, Bi$_2$Se$_3$, and PbTe.
  We think the large anharmonicity is probably due to the Na intercalated layered structures.
 In Na$_2$MgSn and Na$_2$MgPb, there are two layers of Na ions loosely
confined between adjacent Mg-Sn (or Mg-Pb) layers.   The possible rattling modes of
the Na ions could suppress the lattice thermal conductivity, as been found in some cage structure materials, such as
Ba$_8$Ga$_{16}$Ge$_{30}$\cite{cage} and  layered structure materials, such as Na$_x$CoO$_2$. \cite{voneshen}

\begin{table}
	\caption{  Comparison of experimental electric conductivity ($\sigma$), absolute Seebeck coefficient ($S$),
                     total  thermal conductivity ($\kappa$),   and figure of merit $ZT$ of Na$_2$MgSn and Na$_2$MgPb at 300 K
                     with other TE materials.  The values with $\ast$ are theoretical
                     ones  or estimated ones based on the theoretical results.
                     In the table, only SnSe sample is a single crystal. }
	\begin{center}
		\begin{tabular}{c|c|c|c|c}
        \hline\hline
			Material    &   $\sigma$ ($\Omega^{-1}$ cm$^{-1}$ )  &   $S$  ( $\mu$V/K)    & $\kappa$ (W/m$\cdot$K) & $ZT$ \\ \hline
			Na$_2$MgSn \cite{NaMgSn}	&  100 & 390  & 1.343*  & 0.34*   \\  \cline{2-4} \hline
			Na$_2$MgPb \cite{NaMgPb} &  2500 &  22*  & 2.27*  &0.016*  \\  \cline{2-4} \hline
             SnSe \cite{SnSe_nature}      & 1.6(a)/10(b)/10.3(c)  &  542(a)/522(b)/515(c) &  0.46(a)/0.7(b)/0.68(c)     & 0.03(a)/0.12(b)/0.12(c)	 \\  \cline{2-5}  \hline
            SnS \cite{AgSnS}               & 0.001& 400 & 1.25 &$\sim$ 0  	 \\  \cline{2-5}  \hline
            Bi$_2$Te$_3$ \cite{Bi2Te3}   &962 & 226  &1.47 &1.003   \\  \cline{2-5}\hline
           PbTe  \cite{PbTe3}  & 200 & 192  & 1.7  & 0.13	 \\  \cline{2-5}  \hline \hline

	\end{tabular}
	\end{center}
\end{table}

Finally, we can compare some experimental properties of  Na$_2$MgSn and Na$_2$MgPb with other TE
materials, as shown in Table V.  It is found that Na$_2$MgSn has a high Seebeck coefficient at room temperatures,
which is higher than those of Bi$_2$Te$_3$ and PbTe, but lower than those of SnSe and SnS.
It  also has a  good electric conductivity, which is much higher than those of SnSe and SnS,
but lower than those of  Bi$_2$Te$_3$ and PbTe.
Based on these experimental data  and our calculated average total thermal conductivity, we
can estimate that the $ZT$ of Na$_2$MgSn is about 0.34 at 300 K, which is comparable with the
$ZT$ values of  Bi$_2$Te$_3$ and PbTe, but much higher than those of SnSe and SnS at the same temperatures.
We also note that in a polycrystalline sample, the  thermal conductivity should be greatly suppressed usually.
For example, the experimental thermal conductivity of polycrystalline ZrTe$_5$ is only 2.2
W/m$\cdot$K at  room temperatures,\cite{ZT1} which is much lower than the experimental and theoretical values
of single crystalline ZrTe$_5$ (about 8 W/m$\cdot$K at 300 K).\cite{ZT2, ZT3}
Therefore, we expect that the thermal conductivity of polycrystalline Na$_2$MgSn should be much smaller than our
calculated value and its $ZT$  could be possibly much larger than 0.34 at 300 K.
It is also noted that the SnSe and SnS have the best performance at high temperatures (more than 700 K),
while we believe that Na$_2$MgSn would have the highest $ZT$  near the room temperatures due to its small band gap.

On the other hand, although Na$_2$MgPb has an ultra-low lattice thermal conductivity, its total $\kappa_L$
 (about 2.27  W/m$\cdot$K)  is not small due to its metallicity. In experiment,
Yamada \textit{et al}  have found a very small electrical resistivity of polycrystalline Na$_2$MgPb
(only 0.4 m$\Omega$ cm at 300 K).
Combined with our theoretical Seebeck coefficient, we can estimate the $ZT$ of Na$_2$MgPb
is very small, which is only about 0.016 at 300 K, as shown in Table V.
Therefore, metallic  Na$_2$MgPb could not be a good TE material.

\section{ CONCLUSIONS}

We have studied the lattice thermal conductivities  and  thermoelectric properties of 
Na$_2$MgSn and Na$_2$MgPb based on the density functional theory and linearized Boltzmann equation.
Despite of their very low mass density and simple crystal structure, both materials show very
low lattice thermal conductivities, compared with other well-known TE materials.
The  lattice thermal conductivities  along $a$ and $c$ axes of Na$_2$MgSn are 1.75 and 0.80 W/m$\cdot$K
respectively at 300 K,   while they are much lower in  Na$_2$MgPb,
which are 0.51 and 0.31 W/m$\cdot$K  along $a$ and $c$ axes  at the same temperature.
The main reason for the low thermal conductivity is due their large anharmonic effect.

Na$_2$MgPb could not be a good TE material due to its metallicity.
However, we predict that Na$_2$MgSn is a potential  room-temperature TE material with a considerable
large $ZT$ of 0.34 at least.
Since the intermetallic  is a large material family, our work can possibly stimulate
further experimental and theoretical works about the thermoelectric research in simple layered intermetallic compounds.

\textit{Note Added:}  During the preparation of this manuscript, B. Peng \textit{et al.} predicted that
Na$_2$MgPb is a Dirac semi-metal, while Na$_2$MgSn is a trivial indirect semiconductor with
a small band gap of about 0.2 eV. \cite{peng}

\begin{acknowledgments}
We thank Dr. Yang Han for invaluable discussions.
This work is supported by the National Key R\&D Program of China (2016YFA0201104) and
 the National Science Foundation of China (Nos. 91622122 and 11474150).
The use of the computational resources in the High Performance Computing Center of Nanjing University
for this work is also acknowledged.
\end{acknowledgments}

\end{document}